\documentclass[twocolumn,english,amssymb,prl,groupedaddress,notitlepage]{revtex4-2}
\usepackage{amsmath,amssymb,graphicx}
\usepackage{newtxtext}
\usepackage{newtxmath}
\usepackage{multirow}
\usepackage{fancyhdr,bm}
\usepackage{booktabs} 
\usepackage{braket}
\usepackage{titlesec}
\titleformat{\paragraph}
{\normalfont\normalsize\bfseries}{\theparagraph}{1em}{}
\titlespacing*{\paragraph}
{0pt}{3.25ex plus 1ex minus .2ex}{1.5ex plus .2ex}
 
\usepackage{color}
\usepackage[colorlinks=true, pdfstartview=FitV, linkcolor=blue, citecolor=blue, urlcolor=blue]{hyperref}

\nonstopmode

\begin{document}
\title{Majorana Crystal in Rhombohedral Graphene}
\author{Chiho Yoon}
\author{Fan Zhang}
\affiliation{Department of Physics, University of Texas at Dallas, Richardson, Texas 75080, USA}

\begin{abstract}
Recent experiments in rhombohedral graphene report an unusual superconducting phase 
emerging from a spin- and valley-polarized quarter-metal state. 
The prevailing interpretation invokes chiral topological superconductivity, 
but the role of the `Fulde-Ferrell' phase factor due to intra-valley pairing has remained largely unexplored. 
Here we show, via a gauge transformation, that this phase is equivalent to 
an ordinary chiral topological superconductor on the triangular lattice, 
while simultaneously forming an extraordinary Majorana crystal on the dual honeycomb lattice 
reminiscent of the Haldane model.
\end{abstract}
\date{\today} 
\maketitle

\indent\textcolor{blue}{\em Introduction}---%
Superconductivity, magnetism, topology, and fractionalization lie at the heart of condensed matter physics. 
Few-layer graphene with rhombohedral stacking has long been theoretically recognized as fertile ground for 
correlated and topological physics and is now experimentally realized as a highly tunable platform for exploring these phenomena and beyond
\cite{Zhang2010a,Zhang2011a,Weitz2010a, Martin2010a,
Lui2011a, Bao2011a, Zhang2011b, Zou2013a,
Velasco2012a, Freitag2012a, Bao2012a, Shi2020a, Han2024c, Liu2024b, 
Han2024a, Sha2024a, Winterer2024a, Geisenhof2021a, 
Zhou2021a, Barrera2022a, Seiler2022a, Zhou2022a, Yang2024a,
Zhou2021b, Zhang2023a}.
This family of 2D semiconductors provides a rare opportunity to study the coexistence and interplay of all four fundamental themes within a single material. Recently, an unprecedented superconducting phase was observed emerging from a spin- and valley-polarized quarter-metal state, offering a unique setting to unify these themes. While theoretical studies consistently identify this phase as a chiral topological superconductor, the effects of the finite momentum carried by Cooper pairs due to intra-valley pairing have so far been neglected, despite their potential significance
~\cite{Han2025b,Chou2025a,Geier2025a,Kim2025a,Wang2024a,Qin2024a,Yoon2026a,Parra-Martinez2025a,
Bernevig2025a, Gaggioli2025a}.

Pair-density-wave (PDW) superconductivity, characterized by Cooper pairing at finite center-of-mass momentum, 
is a recurring phase in unconventional superconductors~\cite{Agterberg2020a,Casalbuoni2004a,Kinnunen2018a,Fradkin2015a,Kloss2016a}.
Originally proposed by Fulde and Ferrell (FF)~\cite{Fulde1964a} and Larkin and Ovchinnikov (LO)~\cite{Larkin1965a} 
in the context of magnetic-field-induced instabilities, the concept has since broadened beyond its original setting. 
FF-type PDWs, where the Cooper-pair momentum takes a single finite value, have received comparatively little attention. 
They are stabilized only under restrictive conditions that explicitly break both time-reversal and inversion symmetries, 
e.g., strong Zeeman fields or spin-orbit couplings, limiting experimental feasibility. 
Unlike other PDWs, FF states lack spatial amplitude modulation and thus few readily observable signatures. 
Moreover, an FF state can be locally gauge-transformed into a zero-momentum superconductor, 
making it often only weakly distinguishable from uniform superconductivity.

However, the intra-valley chiral PDW state is qualitatively distinct. 
First, it requires neither a Zeeman field nor spin-orbit coupling, 
as the breaking of time-reversal and inversion symmetries is intrinsic to the quarter-metal state. 
Second, the spatial modulation resides in the pairing phase rather than the amplitude. 
Third, unlike conventional cases, the Fermi surface is extremely small whereas the Cooper-pair momentum is extremely large.
Fourth, the Cooper-pair momentum $2K$ (or $2K$') is a special vector invariant under $\mathcal{C}_{3z}$. 
Taken together, these features suggest that this PDW is inherently 2D. 
In particular, the seemingly FF-like phase factor $e^{i2\bm{K}\cdot\bm{R}}$ may instead be interpreted as a symmetric superposition 
$\sum_{j=1,2,3}e^{i2\bm{K}_j\cdot\bm{R}}$, where $K_j$ (j=1,2,3) are the three vectors generated by 
$\mathcal{C}_{3z}$ acting on $K$.
This naturally raises the question of what are the physical implications of such a pairing structure, 
and how the intra-valley chiral PDW differs from both a conventional FF state and a zero-momentum spinless chiral superconductor.

Here we demonstrate that the intra-$K$-valley chiral PDW state is equivalent to 
a zero-momentum topological chiral $p$-wave superconductor on the triangular sublattice to which the electrons are polarized,
with an emergent low-energy Majorana crystal localized on the dual honeycomb sublattice. 
The Peierls phase associated with the momentum shift between the $\Gamma$ and $K$ points, 
corresponding to a $\pm\pi$ flux per triangle, 
is naturally and exactly generated by the vortex-antivortex lattice that hosts the Majorana crystal. 
The resulting Majorana crystal model, reminiscent of the Haldane model, 
has a direct or indirect energy gap and carries a nontrivial Chern number of $\pm 1$.

\indent\textcolor{blue}{\em Topological superconductor}---%
We begin with a description of a chiral $p$-wave topological superconductor on the triangular lattice, 
which we will refer to as the A sublattice.
The corresponding Bogoliubov-de Gennes (BdG) Hamiltonian consists of the chemical potential term, 
the normal state Hamiltonian $\hat{H}_{\rm N}$, and the superconducting order parameter term $\hat{H}_{\Delta}$:
\begin{align}
\begin{split}
\hat{H}_{\rm N} &= \sum_{\bm{R}}\sum_{l =1,2,3}
t_{\bm{R}+\bm{a}_{l},\bm{R}}
e^{\frac{ie}{\hbar c}\int_{\bm{R}}^{\bm{R}+\bm{a}_{l}} \bm{A}\cdot d\bm{l}}
c_{\bm{R}+\bm{a}_{l}}^{\dagger}c_{\bm{R}}
+ \mathrm{H.c.}\,,
\\
\hat{H}_{\Delta} &= \sum_{\bm{R}}\sum_{l =1,2,3}
\Delta_{\bm{R}+\bm{a}_{l},\bm{R}}\,\omega^{l}
c_{\bm{R}+\bm{a}_{l}}^{\dagger}c_{\bm{R}}^{\dagger}
+ \mathrm{H.c.}\,,
\label{eq:H_normal}
\end{split}
\end{align}
where $c_{\bm{R}}^{\dagger}$ is the creation operator of an electron on the triangular A sublattice 
$\bm{R}=n_1\bm{a}_1+n_2\bm{a}_2$ with the primitive lattice vectors $\bm{a}_1=a(1,\,0)$, 
$\bm{a}_{2}=a(-1/2,\,\sqrt{3}/2)$, and $\bm{a}_{3}=-\bm{a}_1-\bm{a}_2$. 
$t_{\bm{R}+\bm{a}_{l},\bm{R}} = -t e^{i\alpha}$ ($t>0$) is the nearest-neighbor hopping, 
which is isotropic and uniform because of the $\mathcal{C}_{3z}$ and translational symmetries.
The Peierls phase encodes the magnetic flux associated with the gauge field $\bm{A}$. 
While $\omega=e^{2\pi i/3}$ encodes the phase winding of the chiral $p$-wave order parameter, 
$\Delta_{\bm{R}+\bm{a}_{l},\bm{R}}$ denotes the center-of-mass superconducting phase, given by
\begin{equation}
\Delta_{\bm{R}+\bm{a}_{l},\bm{R}}
=
\Delta_{0}\,
e^{i\theta(\bm{R})}
\exp\!\left[
\frac{i}{2}\int_{\bm{R}}^{\bm{R}+\bm{a}_{l}}
\nabla\theta\cdot d\bm{l}
\right]\,,
\label{eq:H_BdG_pairing}
\end{equation}
with $\Delta_{0}>0$.
This equation provides a lattice regularization of the superconducting phase 
field $\theta(\bm{r})$~\cite{Melikyan2007a, Murray2015a, Cvetkovic2015a}.
In the absence of magnetic field and (anti)vortex, $\bm{A}=0$ and $\theta=0$. 
In particular, the Cooper pair momentum is zero.

\begin{figure}[t!]
\centering
\includegraphics[width=1.0\linewidth]{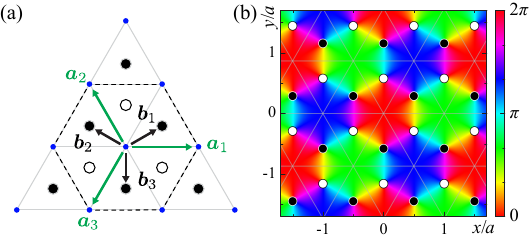}
\caption{(a) Embedded vortices (black dots) and antivortices (open dots) at the centers of the up and down triangles 
of a triangular lattice (blue dots). $\bm{a}_{1}$ and $\bm{a}_{2}$ are the primitive lattice vectors. 
$+\bm{b}_i$ and $-\bm{b}_i$ ($i=1,2,3$) are the positions of vortices and antivortices relative to a lattice point, respectively. 
The blue dashed lines outline a $\sqrt{3}\times\sqrt{3}$ supercell.
(b) Color plot of the superconducting phase field $\theta(\bm{r})$.}
\label{fig1}
\end{figure}

Now embed a vortex-antivortex lattice of vorticity $\pm N$ to the superconductor as follows:
\begin{align}
\nabla\times\nabla\theta(\bm{r}) = 2\pi N \hat{z} \sum_{\bm{R}} \sum_{\tau=\pm}
\tau\delta(\bm{r}-\bm{R}-\tau\bm{b}_{1})\,,
\label{eq:flux}
\end{align}
where $\bm{b}_1\!=\!(2\bm{a}_1+\bm{a}_2)/3$, $\bm{b}_2\!=\!(2\bm{a}_2+\bm{a}_3)/3$, and $\bm{b}_3\!=\!-\bm{b}_1-\bm{b}_2$.
The vortices and antivortices are placed at the centers of the up and down triangles [Fig.~\ref{fig1}(a)], respectively, 
which we hereafter denote as the C and B sublattices. This configuration corresponds to $N$ flux quanta 
(one flux quantum: $\phi_0\!=\!hc/2e$ or $\pi$ magnetic flux) per triangle~\footnote{The delta function, i.e., zero core size, 
is consistent with the low-density and flat-band nature of graphene superconductivity.}, 
equivalent to an effective field of $7.9\times 10^4$~T for the lattice of graphene, 
with the flux sign alternating between the up and down triangles.

\begin{figure*}[t!]
\centering
\includegraphics[width=1.0\linewidth]{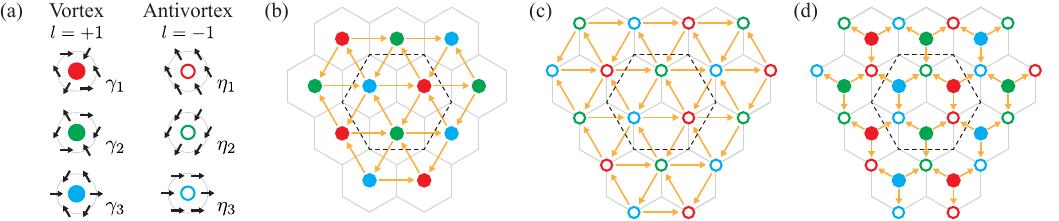}
\caption{
(a) Schematics of the spatial phase-structures of the six types of MBSs,   
localized at vortex (solid circle) and antivortex (open circle) cores. 
Red, green, and blue colors indicate their phases of $2\pi/3$, $4\pi/3$, and 0 along the $+\hat x$ direction, respectively.
Arrows indicate the corresponding phases.
(b-d) Schematics of the the Majorana crystal relative to the graphene honeycomb lattice of the layer to which the electrons are polarized,   
as well as the (b) vortex-vortex, (c) antivortex-antivortex, and (d) vortex-antivortex couplings. 
The dashed lines outline the same $\sqrt{3}\times\sqrt{3}$ supercell shown in Fig~\ref{fig1}(a).}
\label{fig2}
\end{figure*}

The natural solution of the superconducting phase field reads
\begin{align}
\theta(\bm{r}) = N \sum_{\bm{R}} \sum_{\tau=\pm}
\tau\arg(\bm{r}-\bm{R}-\tau\bm{b}_{1})\;\bmod\;{2\pi}\,.
\label{eq:SUM}
\end{align}
However, Eq.~(\ref{eq:SUM}) is only conditionally convergent, because although the monopole moment of each unit cell vanishes, 
higher-order moments such as dipole and quadrupole may remain. This situation is analogous to the convergence issues encountered 
in the Ewald summation of ionic crystals~\cite{Marder2015a}. To guarantee a well-defined $\theta(\bm{r})$, 
we construct the following minimal supercell [Fig.~\ref{fig1}(a)] 
in which the monopole, dipole, and quadrupole moments all vanish, ensuring convergence:
\begin{align}
\theta(\bm{r}) = N \sum_{\bm{R}'} \sum_{l=1,2,3}\sum_{\tau=\pm}
\tau\arg(\bm{r}-\bm{R'}-\tau\bm{b}_{l})\;\bmod\;{2\pi}\,,
\label{eq:theta}
\end{align}
where $\bm{R}' =n_{1}(2\bm{a}_{1}+\bm{a}_{2}) + n_{2}(2\bm{a}_{2}+\bm{a}_{3})$ are the superlattice vectors. 
For a closed triangular path encircling a single (anti)vortex, 
one finds $\oint_{\triangle(\triangledown)}\nabla\theta\cdot d\bm l = \pm2N\pi$. 

The $\mathcal{C}_{3z}$ symmetry further dictates $\int_{\bm{R}}^{\bm{R}+\bm{a}_{l}}\nabla\theta\cdot d\bm{l}=2e/\hbar c\int_{\bm{R}}^{\bm{R}+\bm{a}_{l}}\bm{A}\cdot d\bm{l}=2N\pi/3$~\footnote{We have used $\nabla\theta-(2e/\hbar c)\bm{A}=0$, 
which is valid when the penetration depth $\lambda_L=0$  
or when the supercurrent $j_s=0$ (in equilibrium without superflow or outside a vortex core). 
The $\lambda_L>0$ case will be addressed in {\it Discussion}.}.
Therefore, in the presence of the vortex-antivortex lattice, the Peierls phase in Eq.~(\ref{eq:H_normal}) is $N\pi/3$,
and the order parameter in Eq.~(\ref{eq:H_BdG_pairing}) is 
$\Delta_{\bm{R}+\bm{a}_{l},\bm{R}}=\Delta_0 e^{i\theta(\bm{R})+iN\pi/3}$.

\indent\textcolor{blue}{\em Gauge transformation}---%
Under translations by the primitive lattice vectors $\bm a_i$, $\theta(\bm r)$ is not invariant [Fig.~\ref{fig1}(b)], 
whereas $\nabla\theta(\bm r)$ is invariant, implying that  
$\theta(\bm r+\bm a_i)-\theta(\bm r)=\theta(\bm R+\bm a_i)-\theta(\bm R)=2N\pi/3\bmod 2\pi$. 
It thus follows that under a lattice translation, 
\begin{equation}
\theta(\bm r+\bm R)-\theta(\bm r)=2N\bm K \cdot\bm R\;\bmod\;{2\pi}\,,
\label{eq:superperiodicity}
\end{equation}
where $K=(2\bm G_1 -\bm G_2)/3$. This superperiodicity implies that the shifted phase field
$\tilde{\theta}(\bm{r})=\theta(\bm{r})-2N\bm{K}\cdot\bm{r}$ has a restored lattice translational symmetry:
$\tilde{\theta}(\bm{r})=\tilde{\theta}(\bm{r}+\bm{R})$. This further implies $\tilde{\theta}(\bm{R})=\tilde{\theta}(0)={\theta}_0$, 
which can be chosen to be zero. Nevertheless, the order parameter reads
\begin{align}
\Delta_{\bm{R}+\bm{a}_{l},\bm{R}} 
= \Delta_{0} e^{i(2N\bm{K})\cdot\left(\bm{R}+\frac{\bm{a}_{l}}{2}\right)} e^{i \left(N\pi + \theta_0\right)}\,.
\end{align}
Under the gauge transformation with $\chi(\bm{R})=N\pi + \theta_0$, i.e.,   
\begin{align}
c_{\bm{R}}^{\dagger} \to e^{\frac{i}{2}\chi(\bm{R})} c_{\bm{R}}^{\dagger}\,,\quad
\bm{A} \to \bm{A} - \frac{\hbar c}{2e} \nabla\chi(\bm{R})\,,\nonumber\\
\Delta_{\bm{R}+\bm{a}_{l},\bm{R}} 
\to e^{-\frac{i}{2}\left[\chi(\bm{R})+\chi(\bm{R}+\bm{a}_{l})\right]}\Delta_{\bm{R}+\bm{a}_{l},\bm{R}}\,,
\end{align}
the Hamiltonians in Eq.~(\ref{eq:H_normal}) become
\begin{align}
\begin{split}
\hat{H}_{\rm{N}} &=
\sum_{\bm{R}}\sum_{l=1,2,3} t e^{i\left(\alpha+\frac{N\pi}{3}\right)} c_{\bm{R}+\bm{a}_{l}}^{\dagger}c_{\bm{R}} + \rm H.c.\,, \\
\hat{H}_{\Delta} &=
\sum_{\bm{R}}\sum_{l=1,2,3} \Delta_{0} e^{i(2N\bm{K})\cdot\left(\bm{R}+\frac{\bm{a}_{l}}{2}\right)}\omega^{l} c_{\bm{R}+\bm{a}_{l}}^{\dagger}c^{\dagger}_{\bm{R}} + \rm H.c.\,.
\end{split}
\end{align}

When $N = 3n$~$(n\in\mathbb{Z})$, the first phase factor in $\hat{H}_{\Delta}$ reduces to $1$, 
amounting to the zero-momentum chiral $p$-wave pairing.  When $N = 3n\pm1$~$(n\in\mathbb{Z})$, 
$2N\bm{K}$ becomes $2\bm{K}$~($2\bm{K}'$), and $\hat{H}_{\Delta}$ 
can describe a chiral $p$-wave PDW state as a result of intra-valley pairing within the valley $K$~($K'$). 
Note that the well-defined vorticity is $N \bmod 3$ because of the $C_{3z}$ symmetry.

The normal-state Hamiltonian $\hat{H}_{\rm{N}}$ acquires a Peierls phase of $N\pi/3$ for the electron hopping 
from $\bm{R}$ to $\bm{R}+\bm{a}_{l}$. Hereafter, we focus on the $N=1$ case for simplicity. 
The band energy reads $E(\bm{k},\alpha) = -2t \sum_{l} \cos(\bm{k}\cdot\bm{a}_l+\alpha)$ 
and $E(\bm{k},\alpha+\pi/3)$ without and with the vortex-antivortex lattice.
Independent of the value of $\alpha$, the global band minima of $E(\bm{k},\alpha)$ 
can only occur at $\Gamma$, $K$, and/or $K'$ points (see End Matter). 
For $0 < \alpha < {\pi}/{3}$, the band edge is at $\Gamma$ point originally, whereas it shifts to $K$ point 
because of the $\pi/3$ Peierls phase produced by the vortex-antivortex lattice (see End Matter).

Altogether, we have demonstrated that the spinless, $K$-valley-polarized, low-density, chiral $p$-wave superconductor 
on the triangular lattice is equivalent to a conventional $\Gamma$-valley chiral $p$-wave superconductor on the same lattice  
with an embedded vortex-antivortex lattice of vorticities $\pm1$ on the dual honeycomb lattice. 

\indent\textcolor{blue}{\em Majorana crystal}---%
Because each (anti)vortex binds an unpaired Majorana bound state (MBSs)
in the spinless $\Gamma$-valley chiral $p$-wave superconductor, 
the low-energy quasiparticles of the $K$-valley PDW state 
form a Majorana crystal on the B and C triangular sublattices. 
Now, we construct a minimal two-band model consistent with $\mathcal{C}_{3z}$ symmetry  
and show that the Majorana bands carry a Chern number $\pm 1$.

We first characterize the MBSs created by $\gamma_{\bm S}^{n}$ and $\eta_{\bm S}^{n}$  
at the cores of vortices and antivortices, respectively, in a spinless chiral $p_x+ip_y$ superconductor. 
The core positions are labeled by $\bm S$ and the vorticities by $\ell=\pm 1$. 
The MBSs take the form of 
\begin{align}
\frac{1}{\sqrt{2}}\left\{e^{\frac{i}{2} [\phi_{n} + (\ell+1)\theta]} c_{\bm{S}} + e^{-\frac{i}{2} [\phi_{n} + (\ell+1)\theta]} c_{\bm{S}}^{\dagger}\right\} 
\end{align}
in the continuum limit. Here $c_{\bm{S}}$ is a coherent superposition of all electron annihilation operators, 
the extra unit of winding beyond $\ell$ reflects the angular momentum of a uniform $p_x+ip_y$ pairing field, 
and $\phi_{n}$ denotes the phase along the $+\hat x$ direction.
In general, the relative phase intrinsic to a MBS enclosing a vortex core winds twice faster than $\theta$, 
whereas it remains a constant when enclosing an antivortex core.

For the MBSs on the honeycomb lattice, the continuous rotational symmetry reduces to $\mathcal{C}_{3z}$ symmetry.
Thus, $\phi_{n} = 2\pi n/3$ can be chosen,  
to ensure their $\pm 2\pi/3$ differences dictated by $\mathcal{C}_{3z}$ symmetry.
Figure~\ref{fig2}(a) sketches the spatial phase-structures of the six types of MBSs. 
Under the $\mathcal{C}_{3z}$ rotation around the (anti)vortex core $\bm{S}$, 
$\mathcal{C}_{3z} \gamma_{\bm{S}}^n	\mathcal{C}_{3z}^{-1} = \gamma_{\bm{S}}^{n+1}$ and
$\mathcal{C}_{3z} \eta_{\bm{S}}^n \mathcal{C}_{3z}^{-1} = \eta_{\bm{S}}^n$.
However, the $\mathcal{C}_{3z}$ center can be any lattice site of the A, B, or C sublattice,  
and the $p_x+ip_y$ PDW state as a result of intra-$K$-valley pairing respects the  
$\mathcal{C}_{3z}^{\rm A}$, $\mathcal{C}_{3z}^{\rm B}$, and $\mathcal{C}_{3z}^{\rm C}$ symmetries 
following the $\bar E_K$ irreducible representation~\cite{Yoon2026a}.
In short, while $\hat H_{\rm N}$ is invariant under the $\mathcal{C}_{3z}^{\rm A}$, 
$\mathcal{C}_{3z}^{\rm B}$, and $\mathcal{C}_{3z}^{\rm C}$ operations,
$\hat H_{\Delta}$ acquires a phase factor $\bar{\omega}$, $1$, and $\omega$, respectively. 

This implies that, besides the rotational symmetry $\mathcal{C}_{3z}^{\rm B}$, 
both $\hat H_{\rm N}$ and $\hat H_{\Delta}$ are invariant under the {\it projective} rotational symmetries 
$\tilde{\mathcal{C}}_{3z}^{\rm{A}}={\mathcal{C}}_{3z}^{\rm{A}}\omega^{\tau_z}$ and 
 $\tilde{\mathcal{C}}_{3z}^{\rm{C}}={\mathcal{C}}_{3z}^{\rm{C}}{\bar\omega}^{\tau_z}$ for the BdG Hamiltonian, 
 where $\tau_z$ acts on the Nambu particle-hole space.
Correspondingly, as summarized in Table~\ref{Table1}, the transformations of $\gamma_{\bm S}^{n}$ and $\eta_{\bm S}^{n}$ 
under the projective symmetry operations are modified from those under the normal rotations. 
We note that the MBS configuration shown in Fig.~\ref{fig2} is consistent with these symmetry considerations, 
as well as the superperiodicity specified in Eq.~(\ref{eq:superperiodicity}).
Remarkably, this configuration is in fact the only possibility to embed a single vortex-antivortex pair
per original unit cell that is compatible with the $\bar{E}_{K}$ irreducible representation:
a vortex at the C sublattice site and an antivortex at the B sublattice site. 

We then consider the couplings between MBSs.
For simplicity, we include only three types of nearest-neighbor couplings, 
i.e., vortex-vortex ($vv$), antivortex-antivortex ($aa$), and vortex-antivortex ($va$), which are depicted in Figs.~\ref{fig2}(c-e).
It is straightforward to show that by imposing the $\tilde{\mathcal{C}}_{3z}^{\rm{A}}$, ${\mathcal{C}}_{3z}^{\rm{B}}$, 
$\tilde{\mathcal{C}}_{3z}^{\rm{C}}$, and translational symmetries,  
all the coupling amplitudes are identical within the same type. 
We denote the three types of couplings by $t_{vv}$, $t_{aa}$, and $t_{va} \in \mathbb{R}$, respectively.
This isotropy and uniformity are consistent with the rule of Josephson coupling between two (anti)vortices~\cite{Biswas2013a}. 
\begin{table}[b!]
\setlength{\tabcolsep}{4pt} 
\centering
\caption{Transformation rules of the Majorana operators under the three (projective) $C_{3z}$ rotations, e.g., 
$\tilde{C}_{3z}^{\rm A}\gamma_{\bm S}^{n}\left(\tilde{C}_{3z}^{\rm A}\right)^{-1} = \gamma_{\mathcal{C}_{3z}^{\rm A}\bm S}^{n-1}$.}
\vspace{0.05in}
\begin{tabular}{c c c c}
\hline\hline
\quad Operator\quad\quad & $\tilde{C}_{3z}^{\rm A}$ & $C_{3z}^{\rm B}$ & $\tilde{C}_{3z}^{\rm C}$\\[4pt]
\hline
$\gamma_{\bm{S}}^{n}$  
& \quad$\gamma_{\mathcal{C}_{3z}^{\rm A}\bm{S}}^{n-1}$\quad\quad 
& \quad$\gamma_{\mathcal{C}_{3z}^{\rm B}\bm{S}}^{n+1}$\quad\quad  
& \quad$\gamma_{\mathcal{C}_{3z}^{\rm C}\bm{S}}^{n}$\quad\quad  \\[8pt]
$\eta_{\bm{S}}^{n}$  
& \quad$\eta_{\mathcal{C}_{3z}^{\rm A}\bm{S}}^{n+1}$\quad\quad 
& \quad$\eta_{\mathcal{C}_{3z}^{\rm B}\bm{S}}^{n}$\quad\quad  
& \quad$\eta_{\mathcal{C}_{3z}^{\rm C}\bm{S}}^{n-1}$\quad\quad  \\[8pt]
\hline\hline
\end{tabular}
\label{Table1}
\end{table}

Thus, the low-energy Majorana crystal (MC) model reads
\begin{equation}
\hat{H}_{\rm MC}
= \sum_{\langle ij \rangle}it_{va} \gamma_{i} \eta_{j}
+  \sum_{\langle\!\langle i j \rangle\!\rangle}
\left(it_{vv}\gamma_{i} \gamma_{j} +it_{aa} \eta_{i} \eta_{j} \right)\,,
\label{eq:MC_2band}
\end{equation}
where $\langle ij\rangle$ and $\langle\!\langle ij\rangle\!\rangle$ denote the nearest-neighbor and 
next-nearest-neighbor pairs on the honeycomb lattice formed by the MBSs. 
The order of the two Majorana operators in Eq.~(\ref{eq:MC_2band}) follows the tail-to-head convention in Fig.~\ref{fig2}.

\begin{figure}[t!]
\centering
\includegraphics[width=1.0\linewidth]{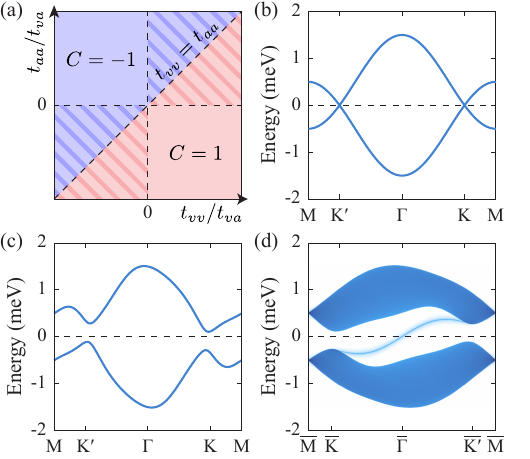}
\caption{(a) The phase diagram of the Majorana crystal model Eq.~(\ref{eq:MC_2band}). 
The Chern number is $1$ ($-1$) for the light red (blue) region. 
There is no indirect gap in the shaded region. 
(b) The Majorana-band Dirac cones at the $K$ and $K$' points 
for the case with {$t_{vv} = t_{aa} = 0$ and $t_{va} = 1$~meV.} 
(c) The bulk and (d) the armchair-edge Majorana band structures for the case with 
{$t_{vv} = 1/5, t_{aa} = -1/15$, and $t_{va} = 1$~meV.}
}
\label{fig3}
\end{figure}

The structure of Eq.~(\ref{eq:MC_2band}) mirrors that of the Haldane model~\cite{Haldane1988a}
with purely imaginary hoppings and vanishing staggered potential. 
Figure~\ref{fig3}(a) shows the phase diagram of this Haldane model of Majoranas. 
Specifically, $t_{va}$ couples the two different sublattices, 
producing the Majorana-band Dirac cones at the $K$ and $K$' points shown in Fig.~\ref{fig3}(b).
Furthermore, $t_v - t_a$ gives rise to the Haldane gap, 
while $t_v + t_a$ provides opposite energy shifts in the two valleys respecting the particle-hole symmetry, 
as exemplified in Fig.~\ref{fig3}(c).

The Majorana crystal model Eq.~(\ref{eq:MC_2band}) reflects the direct vs. indirect gap nature 
and the topological character of the intra-$K$-valley chiral PDW state.
The two Majorana bands are directly gapped as long as {$t_{vv}\neq t_{aa}$ and $t_{va}\neq 0$}, 
and the global gap exists when {$t_{vv}t_{aa}<0$}. 
The Chern number is determined by {$C =\mathrm{sgn}(t_{vv} - t_{aa})$}. 
Figure~\ref{fig3}(d) presents the spectrum of the Majorana crystal 
with an armchair edge in a fully gapped case, obtained using the surface Green's function method. 
Evidently, a single chiral Majorana edge state emerges.

\indent\textcolor{blue}{\em Discussion}---%
In rhombohedral graphene, the large-electric-field-gapped bands are ultra-flat at low electron densities. 
The valley-polarized quarter-metal bands are mainly determined by the emergent order parameter, 
which acts as a Haldane mass at low energy, producing opposite energy shifts in the two valleys. 
Moreover, the low-energy electrons are polarized onto a single triangular sublattice of the outermost layer. 
Consequently, this order parameter can be modeled by $E(\bm{k},\pi/2)$, 
implying that the quarter-metal superconductivity is equivalent to a $\Gamma$-valley chiral $p$-wave 
superconductor on the same sublattice, described by $\alpha=\pi/6$ in Eq.~(\ref{eq:H_normal}), 
while simultaneously forming a Majorana crystal on the other two sublattices.

We have assumed a vanishing penetration depth, $\lambda_L=0$, for simplicity, although this assumption can be relaxed.
For finite $\lambda_L$, the relation $\bm{A} = ({\Phi_{0}}/{2\pi})\nabla\theta$ no longer holds, and the fluxoid quantization,  
including the contribution from the loop current, reads 
\begin{equation}
\oint \bm{A}\cdot d\bm{l}
=\frac{\Phi_{0}}{2\pi}\oint \nabla\theta\cdot d\bm{l}
-\frac{4\pi\lambda_L^2}{c}\oint \bm{j}_s\cdot d\bm{l}\,.
\end{equation}
Consequently, the loop current modifies the $\pi$ magnetic flux per triangle, or equivalently, 
the $\pi/3$ Peierls phase induced by the embedded vortex-antivortex lattice.

From the phase perspective, we have identified six types of MBSs allowed by the symmetries. 
However, their couplings imply only two Majorana bands also consistent with the symmetries. 
Since the phase carries gauge freedom, the phase difference and magnetic flux are more physically meaningful. 
Indeed, the flux given by Eq.~(\ref{eq:flux}) and $\nabla\theta$ from Eq.~(\ref{eq:theta}) preserve 
the original lattice translational symmetry, supporting the two-band structure of our Majorana crystal model.

Our theory and its application to quarter-metal superconductivity in rhombohedral graphene 
rely on the $\mathcal{C}_{3z}$ symmetry of the parent phase, 
which pins vortices and antivortices to distinct triangular sublattices. 
Strong Fermi-surface anisotropy instead favors nematic order, whose breaking of $\mathcal{C}_{3z}$ symmetry 
unpins vortices and antivortices, allowing their annihilation and stabilizing a FF state with 1D phase modulation.
Recent experiments on rhombohedral hexalayer graphene report a nematic transition 
in the quarter-metal phase~\cite{Nguyen2025a,Qin2026a}, which may provide such an example. 
Starting instead from a quarter metal with an anisotropic Fermi surface but without nematic order, 
a recent macroscopic theory of superconductivity~\cite{Gaggioli2025a, Christos2025a, Gil2025a} ingeniously shows 
the possible emergence of a vortex-antivortex lattice, forming a $\mathcal{C}_{3z}$-invariant pattern similar to ours. 
However, that pattern is incommensurate with the underlying lattice, 
and the emergent lattice constant is several orders of magnitude larger. Furthermore, 
it neglects the $e^{i2\bm{K}\cdot\bm{R}}$ factor, which we find essential.

\indent\textcolor{blue}{\em Acknowledgement}---%
This work was supported by the National Science Foundation under Grants 
No. DMR-1945351, No. DMR-2324033, and No. DMR-2414726 
and by the Welch Foundation under Grant No. AT-2264-20250403.

\bibliographystyle{apsrev4-2_edit_200521}
\bibliography{MC_ref}{}

@misc{Gil2025a,
	archiveprefix = {arXiv},
	author = {Avigail Gil and Erez Berg},
	eprint = {2504.19321},
	title = {Charge and pair density waves in a spin and valley-polarized system at a Van-Hove singularity},
	url = {https://arxiv.org/abs/2504.19321},
	year = {2025}}

@article{Geier2025a,
	author = {Geier, Max and Davydova, Margarita and Fu, Liang},
	date = {2025/12/01},
	doi = {10.1038/s41467-025-66902-6},
	id = {Geier2025},
	isbn = {2041-1723},
	journal = {Nat. Commun.},
	number = {1},
	pages = {232},
	title = {Chiral and topological superconductivity in isospin polarized multilayer graphene},
	url = {https://doi.org/10.1038/s41467-025-66902-6},
	volume = {17},
	year = {2025}}

@misc{Bernevig2025a,
	archiveprefix = {arXiv},
	author = {B. Andrei Bernevig and Yves H. Kwan},
	eprint = {2503.09692},
	title = {"Berry Trashcan" Model of Interacting Electrons in Rhombohedral Graphene},
	url = {https://arxiv.org/abs/2503.09692},
	year = {2025}}

@article{Zhang2010a,
	author = {Zhang, Fan and Sahu, Bhagawan and Min, Hongki and MacDonald, A. H.},
	doi = {10.1103/physrevb.82.035409},
	issn = {1098-0121},
	journal = {Phys. Rev. B},
	number = {3},
	pages = {035409},
	title = {{Band structure of ABC-stacked graphene trilayers}},
	volume = {82},
	year = {2010}}

@article{Winterer2024a,
	author = {Winterer, Felix and Geisenhof, Fabian R. and Fernandez, Noelia and Seiler, Anna M. and Zhang, Fan and Weitz, R. Thomas},
	doi = {10.1038/s41567-023-02327-6},
	issn = {1745-2473},
	journal = {Nat. Phys.},
	number = {3},
	pages = {422--427},
	title = {{Ferroelectric and spontaneous quantum Hall states in intrinsic rhombohedral trilayer graphene}},
	volume = {20},
	year = {2024}}

@article{Geisenhof2021a,
	author = {Geisenhof, Fabian R. and Winterer, Felix and Seiler, Anna M. and Lenz, Jakob and Xu, Tianyi and Zhang, Fan and Weitz, R. Thomas},
	doi = {10.1038/s41586-021-03849-w},
	issn = {0028-0836},
	journal = {Nature},
	number = {7879},
	pages = {53--58},
	pmid = {34616059},
	title = {{Quantum anomalous Hall octet driven by orbital magnetism in bilayer graphene}},
	volume = {598},
	year = {2021}}

@article{Velasco2012a,
	author = {Velasco, J. and Jing, L. and Bao, W. and Lee, Y. and Kratz, P. and Aji, V. and Bockrath, M. and Lau, C. N. and Varma, C. and Stillwell, R. and Smirnov, D. and Zhang, Fan and Jung, J. and MacDonald, A. H.},
	doi = {10.1038/nnano.2011.251},
	issn = {1748-3387},
	journal = {Nat. Nanotechnol.},
	number = {3},
	pages = {156--160},
	pmid = {22266634},
	title = {{Transport spectroscopy of symmetry-broken insulating states in bilayer graphene}},
	volume = {7},
	year = {2012}}

@article{Zhang2011a,
	author = {Zhang, Fan and Jung, Jeil and Fiete, Gregory A. and Niu, Qian and MacDonald, Allan H.},
	doi = {10.1103/physrevlett.106.156801},
	issn = {0031-9007},
	journal = {Phys. Rev. Lett.},
	number = {15},
	pages = {156801},
	pmid = {21568592},
	title = {{Spontaneous Quantum Hall States in Chirally Stacked Few-Layer Graphene Systems}},
	volume = {106},
	year = {2011}}

@article{Han2024c,
	author = {Han, Tonghang and Lu, Zhengguang and Scuri, Giovanni and Sung, Jiho and Wang, Jue and Han, Tianyi and Watanabe, Kenji and Taniguchi, Takashi and Park, Hongkun and Ju, Long},
	date = {2024/02/01},
	doi = {10.1038/s41565-023-01520-1},
	id = {Han2024},
	isbn = {1748-3395},
	journal = {Nat. Nanotechnol.},
	number = {2},
	pages = {181--187},
	title = {Correlated insulator and Chern insulators in pentalayer rhombohedral-stacked graphene},
	url = {https://doi.org/10.1038/s41565-023-01520-1},
	volume = {19},
	year = {2024}}

@article{Weitz2010a,
	author = {Weitz, R. T. and Allen, M. T. and Feldman, B. E. and Martin, J. and Yacoby, A.},
	date = {2010/11/05},
	doi = {10.1126/science.1194988},
	journal = {Science},
	journal1 = {Science},
	journal2 = {Science},
	month = {2025/02/08},
	n2 = {Bilayer graphene samples are expected exhibit quantum Hall states that are ferromagnetic with different types of spin ordering. Weitz et al. (p. 812, published online 14 October) studied the conductance of high-quality suspended bilayer graphene samples. They used an applied perpendicular electric field to induce transitions between the different broken-symmetry states that appear at low carrier densities and deduced their order parameters. These states appeared in both the absence of a magnetic field, as well as in the presence of a symmetry-breaking magnetic field. They also showed that, even in absence of both an applied magnetic or electric field, the bilayer exhibits an energy gap, which indicates that electron-electron interactions contribute to the band structure. Ferromagnetic quantum Hall states were observed in suspended bilayer graphene samples in both zero and high magnetic fields. The single-particle energy spectra of graphene and its bilayer counterpart exhibit multiple degeneracies that arise through inherent symmetries. Interactions among charge carriers should spontaneously break these symmetries and lead to ordered states that exhibit energy gaps. In the quantum Hall regime, these states are predicted to be ferromagnetic in nature, whereby the system becomes spin polarized, layer polarized, or both. The parabolic dispersion of bilayer graphene makes it susceptible to interaction-induced symmetry breaking even at zero magnetic field. We investigated the underlying order of the various broken-symmetry states in bilayer graphene suspended between top and bottom gate electrodes. We deduced the order parameter of the various quantum Hall ferromagnetic states by controllably breaking the spin and sublattice symmetries. At small carrier density, we identified three distinct broken-symmetry states, one of which is consistent with either spontaneously broken time-reversal symmetry or spontaneously broken rotational symmetry.},
	number = {6005},
	pages = {812--816},
	publisher = {American Association for the Advancement of Science},
	title = {Broken-Symmetry States in Doubly Gated Suspended Bilayer Graphene},
	type = {doi: 10.1126/science.1194988},
	url = {https://doi.org/10.1126/science.1194988},
	volume = {330},
	year = {2010},
	year1 = {2010}}

@article{Martin2010a,
	author = {Martin, J. and Feldman, B. E. and Weitz, R. T. and Allen, M. T. and Yacoby, A.},
	date = {2010/12/15/},
	day = {15},
	doi = {10.1103/PhysRevLett.105.256806},
	id = {10.1103/PhysRevLett.105.256806},
	j1 = {PRL},
	journal = {Phys. Rev. Lett.},
	journal1 = {Phys. Rev. Lett.},
	month = {12},
	number = {25},
	pages = {256806},
	publisher = {American Physical Society},
	title = {Local Compressibility Measurements of Correlated States in Suspended Bilayer Graphene},
	url = {https://link.aps.org/doi/10.1103/PhysRevLett.105.256806},
	volume = {105},
	year = {2010}}

@article{Lui2011a,
	author = {Lui, Chun Hung and Li, Zhiqiang and Mak, Kin Fai and Cappelluti, Emmanuele and Heinz, Tony F.},
	date = {2011/12/01},
	doi = {10.1038/nphys2102},
	id = {Lui2011},
	isbn = {1745-2481},
	journal = {Nat. Phys.},
	number = {12},
	pages = {944--947},
	title = {Observation of an electrically tunable band gap in trilayer graphene},
	url = {https://doi.org/10.1038/nphys2102},
	volume = {7},
	year = {2011}}

@article{Bao2011a,
	author = {Bao, W. and Jing, L. and Velasco, J. and Lee, Y. and Liu, G. and Tran, D. and Standley, B. and Aykol, M. and Cronin, S. B. and Smirnov, D. and Koshino, M. and McCann, E. and Bockrath, M. and Lau, C. N.},
	date = {2011/12/01},
	doi = {10.1038/nphys2103},
	id = {Bao2011},
	isbn = {1745-2481},
	journal = {Nat. Phys.},
	number = {12},
	pages = {948--952},
	title = {Stacking-dependent band gap and quantum transport in trilayer graphene},
	url = {https://doi.org/10.1038/nphys2103},
	volume = {7},
	year = {2011}}

@article{Zhang2011b,
	author = {Zhang, Liyuan and Zhang, Yan and Camacho, Jorge and Khodas, Maxim and Zaliznyak, Igor},
	date = {2011/12/01},
	doi = {10.1038/nphys2104},
	id = {Zhang2011},
	isbn = {1745-2481},
	journal = {Nat. Phys.},
	number = {12},
	pages = {953--957},
	title = {The experimental observation of quantum Hall effect of l=3 chiral quasiparticles in trilayer graphene},
	url = {https://doi.org/10.1038/nphys2104},
	volume = {7},
	year = {2011}}

@article{Zou2013a,
	author = {Zou, K. and Zhang, Fan and Clapp, C. and MacDonald, A. H. and Zhu, J.},
	date = {2013/02/13},
	doi = {10.1021/nl303375a},
	isbn = {1530-6984},
	journal = {Nano Lett.},
	journal1 = {Nano Letters},
	journal2 = {Nano Lett.},
	month = {02},
	number = {2},
	pages = {369--373},
	publisher = {American Chemical Society},
	title = {Transport Studies of Dual-Gated ABC and ABA Trilayer Graphene: Band Gap Opening and Band Structure Tuning in Very Large Perpendicular Electric Fields},
	type = {doi: 10.1021/nl303375a},
	url = {https://doi.org/10.1021/nl303375a},
	volume = {13},
	year = {2013},
	year1 = {2013}}

@article{Freitag2012a,
	author = {Freitag, F. and Trbovic, J. and Weiss, M. and Sch{\"o}nenberger, C.},
	date = {2012/02/13/},
	day = {13},
	doi = {10.1103/PhysRevLett.108.076602},
	id = {10.1103/PhysRevLett.108.076602},
	j1 = {PRL},
	journal = {Phys. Rev. Lett.},
	journal1 = {Phys. Rev. Lett.},
	month = {02},
	number = {7},
	pages = {076602},
	publisher = {American Physical Society},
	title = {Spontaneously Gapped Ground State in Suspended Bilayer Graphene},
	url = {https://link.aps.org/doi/10.1103/PhysRevLett.108.076602},
	volume = {108},
	year = {2012}}

@article{Bao2012a,
	author = {Bao, Wenzhong and Velasco, Jairo and Zhang, Fan and Jing, Lei and Standley, Brian and Smirnov, Dmitry and Bockrath, Marc and MacDonald, Allan H. and Lau, Chun Ning},
	date = {2012/07/03},
	doi = {10.1073/pnas.1205978109},
	journal = {Proc. Natl. Acad. Sci. U.S.A.},
	journal1 = {Proc. Natl. Acad. Sci. U.S.A.},
	journal2 = {Proc. Natl. Acad. Sci. U.S.A.},
	month = {2025/02/08},
	n2 = {At the charge neutrality point, bilayer graphene (BLG) is strongly susceptible to electronic interactions and is expected to undergo a phase transition to a state with spontaneously broken symmetries. By systematically investigating a large number of single-and double-gated BLG devices, we observe a bimodal distribution of minimum conductivities at the charge neutrality point. Although σmin is often approximately 2?3 e2/h (where e is the electron charge and h is Planck?s constant), it is several orders of magnitude smaller in BLG devices that have both high mobility and low extrinsic doping. The insulating state in the latter samples appears below a transition temperature Tc of approximately 5 K and has a T = 0 energy gap of approximately 3 meV. Transitions between these different states can be tuned by adjusting disorder or carrier density.},
	number = {27},
	pages = {10802--10805},
	publisher = {Proc. Natl. Acad. Sci. U.S.A.},
	title = {Evidence for a spontaneous gapped state in ultraclean bilayer graphene},
	type = {doi: 10.1073/pnas.1205978109},
	url = {https://doi.org/10.1073/pnas.1205978109},
	volume = {109},
	year = {2012},
	year1 = {2012}}

@article{Shi2020a,
	author = {Shi, Yanmeng and Xu, Shuigang and Yang, Yaping and Slizovskiy, Sergey and Morozov, Sergey V. and Son, Seok-Kyun and Ozdemir, Servet and Mullan, Ciaran and Barrier, Julien and Yin, Jun and Berdyugin, Alexey I. and Piot, Benjamin A. and Taniguchi, Takashi and Watanabe, Kenji and Fal'ko, Vladimir I. and Novoselov, Kostya S. and Geim, A. K. and Mishchenko, Artem},
	date = {2020/08/01},
	doi = {10.1038/s41586-020-2568-2},
	id = {Shi2020},
	isbn = {1476-4687},
	journal = {Nature},
	number = {7820},
	pages = {210--214},
	title = {Electronic phase separation in multilayer rhombohedral graphite},
	url = {https://doi.org/10.1038/s41586-020-2568-2},
	volume = {584},
	year = {2020}}

@article{Sha2024a,
	author = {Sha, Yating and Zheng, Jian and Liu, Kai and Du, Hong and Watanabe, Kenji and Taniguchi, Takashi and Jia, Jinfeng and Shi, Zhiwen and Zhong, Ruidan and Chen, Guorui},
	date = {2024/04/26},
	doi = {10.1126/science.adj8272},
	journal = {Science},
	journal1 = {Science},
	journal2 = {Science},
	month = {2025/02/08},
	n2 = {Degeneracies in multilayer graphene, including spin, valley, and layer degrees of freedom, can be lifted by Coulomb interactions, resulting in rich broken-symmetry states. Here, we report a ferromagnetic state in charge-neutral ABCA-tetralayer graphene driven by proximity-induced spin-orbit coupling from adjacent tungsten diselenide. The ferromagnetic state is identified as a Chern insulator with a Chern number of 4; its maximum Hall resistance reaches 78{\%} quantization at zero magnetic field and is fully quantized at either 0.4 or ?1.5 tesla. Three distinct broken-symmetry insulating states, layer-antiferromagnet, Chern insulator, and layer-polarized insulator, along with their transitions, can be continuously tuned by the vertical displacement field. In this system, the magnetic order of the Chern insulator can be switched by three knobs, including magnetic field, electrical doping, and vertical displacement field. Properties of two-dimensional layered materials depend sensitively on the stacking arrangement of the layers. Rhombohedral multilayer graphene has been predicted to host topologically nontrivial states such as the Chern insulator. Sha et al. observed such a state in four-layer rhombohedral graphene encapsulated by hexagonal boron nitride and placed next to a layer of tungsten diselenide. The role of this last layer was to induce spin-orbit coupling, which is normally negligible in graphene. Transport measurements indicated ferromagnetism and a quantization of Hall resistance stabilized by magnetic fields, corresponding to a Chern number of 4. ?Jelena Stajic},
	number = {6694},
	pages = {414--419},
	publisher = {American Association for the Advancement of Science},
	title = {Observation of a Chern insulator in crystalline ABCA-tetralayer graphene with spin-orbit coupling},
	type = {doi: 10.1126/science.adj8272},
	url = {https://doi.org/10.1126/science.adj8272},
	volume = {384},
	year = {2024},
	year1 = {2024}}

@article{Barrera2022a,
	author = {de la Barrera, Sergio C. and Aronson, Samuel and Zheng, Zhiren and Watanabe, Kenji and Taniguchi, Takashi and Ma, Qiong and Jarillo-Herrero, Pablo and Ashoori, Raymond},
	date = {2022/07/01},
	doi = {10.1038/s41567-022-01616-w},
	id = {de la Barrera2022},
	isbn = {1745-2481},
	journal = {Nat. Phys.},
	number = {7},
	pages = {771--775},
	title = {Cascade of isospin phase transitions in Bernal-stacked bilayer graphene at zero magnetic field},
	url = {https://doi.org/10.1038/s41567-022-01616-w},
	volume = {18},
	year = {2022}}

@misc{Yang2024a,
	archiveprefix = {arXiv},
	author = {Jixiang Yang and Xiaoyan Shi and Shenyong Ye and Chiho Yoon and Zhengguang Lu and Vivek Kakani and Tonghang Han and Junseok Seo and Lihan Shi and Kenji Watanabe and Takashi Taniguchi and Fan Zhang and Long Ju},
	eprint = {2408.09906},
	title = {Impact of Spin-Orbit Coupling on Superconductivity in Rhombohedral Graphene},
	url = {https://arxiv.org/abs/2408.09906},
	year = {2025}}

@article{Zhang2023a,
	author = {Zhang, Yiran and Polski, Robert and Thomson, Alex and Lantagne-Hurtubise, {\'E}tienne and Lewandowski, Cyprian and Zhou, Haoxin and Watanabe, Kenji and Taniguchi, Takashi and Alicea, Jason and Nadj-Perge, Stevan},
	doi = {10.1038/s41586-022-05446-x},
	issn = {0028-0836},
	journal = {Nature},
	number = {7943},
	pages = {268--273},
	pmid = {36631645},
	title = {{Enhanced superconductivity in spin--orbit proximitized bilayer graphene}},
	volume = {613},
	year = {2023}}

@article{Zhou2021a,
	author = {Zhou, Haoxin and Xie, Tian and Ghazaryan, Areg and Holder, Tobias and Ehrets, James R. and Spanton, Eric M. and Taniguchi, Takashi and Watanabe, Kenji and Berg, Erez and Serbyn, Maksym and Young, Andrea F.},
	doi = {10.1038/s41586-021-03938-w},
	issn = {0028-0836},
	journal = {Nature},
	number = {7881},
	pages = {429--433},
	pmid = {34469943},
	title = {{Half- and quarter-metals in rhombohedral trilayer graphene}},
	volume = {598},
	year = {2021}}

@article{Zhou2022a,
	author = {Zhou, Haoxin and Holleis, Ludwig and Saito, Yu and Cohen, Liam and Huynh, William and Patterson, Caitlin L. and Yang, Fangyuan and Taniguchi, Takashi and Watanabe, Kenji and Young, Andrea F.},
	doi = {10.1126/science.abm8386},
	issn = {0036-8075},
	journal = {Science},
	number = {6582},
	pages = {774--778},
	pmid = {35025604},
	title = {{Isospin magnetism and spin-polarized superconductivity in Bernal bilayer graphene}},
	volume = {375},
	year = {2022}}

@article{Han2024a,
	author = {Han, Tonghang and Lu, Zhengguang and Yao, Yuxuan and Yang, Jixiang and Seo, Junseok and Yoon, Chiho and Watanabe, Kenji and Taniguchi, Takashi and Fu, Liang and Zhang, Fan and Ju, Long},
	doi = {10.1126/science.adk9749},
	issn = {0036-8075},
	journal = {Science},
	number = {6696},
	pages = {647--651},
	pmid = {38723084},
	title = {{Large quantum anomalous Hall effect in spin-orbit proximitized rhombohedral graphene}},
	volume = {384},
	year = {2024}}

@article{Seiler2022a,
	author = {Seiler, Anna M. and Geisenhof, Fabian R. and Winterer, Felix and Watanabe, Kenji and Taniguchi, Takashi and Xu, Tianyi and Zhang, Fan and Weitz, R. Thomas},
	doi = {10.1038/s41586-022-04937-1},
	issn = {0028-0836},
	journal = {Nature},
	number = {7922},
	pages = {298--302},
	pmid = {35948716},
	title = {{Quantum cascade of correlated phases in trigonally warped bilayer graphene}},
	volume = {608},
	year = {2022}}

@article{Liu2024b,
	author = {Liu, Kai and Zheng, Jian and Sha, Yating and Lyu, Bosai and Li, Fengping and Park, Youngju and Ren, Yulu and Watanabe, Kenji and Taniguchi, Takashi and Jia, Jinfeng and Luo, Weidong and Shi, Zhiwen and Jung, Jeil and Chen, Guorui},
	doi = {10.1038/s41565-023-01558-1},
	issn = {1748-3387},
	journal = {Nat. Nanotechnol.},
	number = {2},
	pages = {188--195},
	pmid = {37996516},
	title = {{Spontaneous broken-symmetry insulator and metals in tetralayer rhombohedral graphene}},
	volume = {19},
	year = {2024}}

@article{Zhou2021b,
	author = {Zhou, Haoxin and Xie, Tian and Taniguchi, Takashi and Watanabe, Kenji and Young, Andrea F.},
	doi = {10.1038/s41586-021-03926-0},
	issn = {0028-0836},
	journal = {Nature},
	number = {7881},
	pages = {434--438},
	pmid = {34469942},
	title = {{Superconductivity in rhombohedral trilayer graphene}},
	volume = {598},
	year = {2021}}

@misc{Qin2026a,
	archiveprefix = {arXiv},
	author = {Peiyu Qin and Hai-Tian Wu and Ron Q. Nguyen and Erin Morissette and Naiyuan J. Zhang and K. Watanabe and T. Taniguchi and J. I. A. Li},
	eprint = {2504.05129},
	title = {Extreme Anisotropy in the Metallic and Superconducting Phases of Rhombohedral Hexalayer Graphene},
	url = {https://arxiv.org/abs/2504.05129},
	year = {2026}}

@misc{Nguyen2025a,
	archiveprefix = {arXiv},
	author = {Ron Q. Nguyen and Hai-Tian Wu and Erin Morissette and Naiyuan J. Zhang and Peiyu Qin and Kenji Watanabe and Takashi Taniguchi and Aaron W. Hui and Dima E. Feldman and J. I. A. Li},
	eprint = {2507.22026},
	title = {A Hierarchy of Superconductivity and Topological Charge Density Wave States in Rhombohedral Graphene},
	url = {https://arxiv.org/abs/2507.22026},
	year = {2025}}

@article{Yoon2026a,
	author = {Yoon, Chiho and Xu, Tianyi and Barlas, Yafis and Zhang, Fan},
	doi = {10.1103/fcdc-9lm3},
	issue = {2},
	journal = {Phys. Rev. Lett.},
	month = {Jan},
	numpages = {9},
	pages = {026603},
	publisher = {American Physical Society},
	title = {Quarter-Metal Superconductivity in Rhombohedral Graphene},
	url = {https://link.aps.org/doi/10.1103/fcdc-9lm3},
	volume = {136},
	year = {2026}}

@article{Cvetkovic2015a,
	author = {Cvetkovic, Vladimir and Vafek, Oskar},
	date = {2015/03/11},
	doi = {10.1038/ncomms7518},
	id = {Cvetkovic2015},
	isbn = {2041-1723},
	journal = {Nat. Commun.},
	number = {1},
	pages = {6518},
	title = {Berry phases and the intrinsic thermal Hall effect in high-temperature cuprate superconductors},
	url = {https://doi.org/10.1038/ncomms7518},
	volume = {6},
	year = {2015}}

@article{Murray2015a,
	author = {Murray, James M. and Vafek, Oskar},
	doi = {10.1103/PhysRevB.92.134520},
	issue = {13},
	journal = {Phys. Rev. B},
	month = {Oct},
	numpages = {7},
	pages = {134520},
	publisher = {American Physical Society},
	title = {Majorana bands, Berry curvature, and thermal Hall conductivity in the vortex state of a chiral $p$-wave superconductor},
	url = {https://link.aps.org/doi/10.1103/PhysRevB.92.134520},
	volume = {92},
	year = {2015}}

@article{Biswas2013a,
	author = {Biswas, Rudro R.},
	doi = {10.1103/PhysRevLett.111.136401},
	issue = {13},
	journal = {Phys. Rev. Lett.},
	month = {Sep},
	numpages = {5},
	pages = {136401},
	publisher = {American Physical Society},
	title = {Majorana Fermions in Vortex Lattices},
	url = {https://link.aps.org/doi/10.1103/PhysRevLett.111.136401},
	volume = {111},
	year = {2013}}

@article{Melikyan2007a,
	author = {Melikyan, Ashot and Te\ifmmode \check{s}\else \v{s}\fi{}anovi\ifmmode \acute{c}\else \'{c}\fi{}, Zlatko},
	doi = {10.1103/PhysRevB.76.094509},
	issue = {9},
	journal = {Phys. Rev. B},
	month = {Sep},
	numpages = {30},
	pages = {094509},
	publisher = {American Physical Society},
	title = {Dirac-Bogoliubov-deGennes quasiparticles in a vortex lattice},
	url = {https://link.aps.org/doi/10.1103/PhysRevB.76.094509},
	volume = {76},
	year = {2007}}

@book{Marder2015a,
	author = {M. P. Marder},
	edition = {2nd},
	publisher = {Wiley},
	title = {Condensed Matter Physics},
	year = {2015}}

@article{Haldane1988a,
	author = {Haldane, F. D. M.},
	doi = {10.1103/PhysRevLett.61.2015},
	issue = {18},
	journal = {Phys. Rev. Lett.},
	month = {Oct},
	numpages = {0},
	pages = {2015--2018},
	publisher = {American Physical Society},
	title = {Model for a Quantum Hall Effect without Landau Levels: Condensed-Matter Realization of the "Parity Anomaly"},
	url = {https://link.aps.org/doi/10.1103/PhysRevLett.61.2015},
	volume = {61},
	year = {1988}}

@article{Agterberg2020a,
	author = {Agterberg, Daniel F. and Davis, J.C. S{\'e}amus and Edkins, Stephen D. and Fradkin, Eduardo and Van Harlingen, Dale J. and Kivelson, Steven A. and Lee, Patrick A. and Radzihovsky, Leo and Tranquada, John M. and Wang, Yuxuan},
	doi = {https://doi.org/10.1146/annurev-conmatphys-031119-050711},
	issn = {1947-5462},
	journal = {Annu. Rev. Condens. Matter Phys.},
	number = {Volume 11, 2020},
	pages = {231-270},
	publisher = {Annual Reviews},
	title = {The Physics of Pair-Density Waves: Cuprate Superconductors and Beyond},
	type = {Journal Article},
	url = {https://www.annualreviews.org/content/journals/10.1146/annurev-conmatphys-031119-050711},
	volume = {11},
	year = {2020}}

@misc{Christos2025a,
	archiveprefix = {arXiv},
	author = {Maine Christos and Pietro M. Bonetti and Mathias S. Scheurer},
	eprint = {2503.15471},
	title = {Finite-momentum pairing and superlattice superconductivity in valley-imbalanced rhombohedral graphene},
	url = {https://arxiv.org/abs/2503.15471},
	year = {2025}}

@article{Gaggioli2025a,
	author = {Gaggioli, Filippo and Guerci, Daniele and Fu, Liang},
	doi = {10.1103/k8sb-rqxf},
	issue = {11},
	journal = {Phys. Rev. Lett.},
	month = {Sep},
	numpages = {7},
	pages = {116001},
	publisher = {American Physical Society},
	title = {Spontaneous Vortex-Antivortex Lattice and Majorana Fermions in Rhombohedral Graphene},
	url = {https://link.aps.org/doi/10.1103/k8sb-rqxf},
	volume = {135},
	year = {2025}}

@article{Parra-Martinez2025a,
	author = {Parra-Mart\'{\i}nez, Guillermo and Jimeno-Pozo, Alejandro and Phong, V\~o Tiến and Sainz-Cruz, H\'ector and Kaplan, Daniel and Emanuel, Peleg and Oreg, Yuval and Pantale\'on, Pierre A. and Silva-Guill\'en, Jos\'e \'Angel and Guinea, Francisco},
	doi = {10.1103/zfmh-rjzc},
	issue = {13},
	journal = {Phys. Rev. Lett.},
	month = {Sep},
	numpages = {9},
	pages = {136503},
	publisher = {American Physical Society},
	title = {Band Renormalization, Quarter Metals, and Chiral Superconductivity in Rhombohedral Tetralayer Graphene},
	url = {https://link.aps.org/doi/10.1103/zfmh-rjzc},
	volume = {135},
	year = {2025}}

@article{Kim2025a,
	author = {Kim, Minho and Timmel, Abigail and Ju, Long and Wen, Xiao-Gang},
	doi = {10.1103/PhysRevB.111.014508},
	issue = {1},
	journal = {Phys. Rev. B},
	month = {Jan},
	numpages = {20},
	pages = {014508},
	publisher = {American Physical Society},
	title = {Topological chiral superconductivity beyond pairing in a Fermi liquid},
	url = {https://link.aps.org/doi/10.1103/PhysRevB.111.014508},
	volume = {111},
	year = {2025}}

@article{Chou2025a,
	author = {Chou, Yang-Zhi and Zhu, Jihang and Das Sarma, Sankar},
	doi = {10.1103/PhysRevB.111.174523},
	issue = {17},
	journal = {Phys. Rev. B},
	month = {May},
	numpages = {13},
	pages = {174523},
	publisher = {American Physical Society},
	title = {Intravalley spin-polarized superconductivity in rhombohedral tetralayer graphene},
	url = {https://link.aps.org/doi/10.1103/PhysRevB.111.174523},
	volume = {111},
	year = {2025}}

@misc{Wang2024a,
	archiveprefix = {arXiv},
	author = {Yan-Qi Wang and Zhi-Qiang Gao and Hui Yang},
	eprint = {2410.05384},
	title = {Chiral superconductivity from parent Chern band and its non-Abelian generalization},
	url = {https://arxiv.org/abs/2410.05384},
	year = {2024}}

@misc{Qin2024a,
	archiveprefix = {arXiv},
	author = {Qiong Qin and Congjun Wu},
	eprint = {2412.07145},
	title = {Chiral finite-momentum superconductivity in the tetralayer graphene},
	url = {https://arxiv.org/abs/2412.07145},
	year = {2024}}

@article{Han2025b,
	author = {Han, Tonghang and Lu, Zhengguang and Hadjri, Zach and Shi, Lihan and Wu, Zhenghan and Xu, Wei and Yao, Yuxuan and Cotten, Armel A. and Sharifi Sedeh, Omid and Weldeyesus, Henok and Yang, Jixiang and Seo, Junseok and Ye, Shenyong and Zhou, Muyang and Liu, Haoyang and Shi, Gang and Hua, Zhenqi and Watanabe, Kenji and Taniguchi, Takashi and Xiong, Peng and Zumb{\"u}hl, Dominik M. and Fu, Liang and Ju, Long},
	date = {2025/07/01},
	doi = {10.1038/s41586-025-09169-7},
	id = {Han2025},
	isbn = {1476-4687},
	journal = {Nature},
	number = {8072},
	pages = {654--661},
	title = {Signatures of chiral superconductivity in rhombohedral graphene},
	url = {https://doi.org/10.1038/s41586-025-09169-7},
	volume = {643},
	year = {2025}}

@article{Kloss2016a,
	author = {Kloss, T and Montiel, X and de Carvalho, V S and Freire, H and P{\'e}pin, C},
	doi = {10.1088/0034-4885/79/8/084507},
	journal = {Rep. Prog. Phys.},
	month = {jul},
	number = {8},
	pages = {084507},
	publisher = {IOP Publishing},
	title = {Charge orders, magnetism and pairings in the cuprate superconductors},
	url = {https://doi.org/10.1088/0034-4885/79/8/084507},
	volume = {79},
	year = {2016}}

@article{Larkin1965a,
	author = {Larkin, A. I. and Ovchinnikov, Y. N.},
	issn = {0031-},
	journal = {Sov. Phys. JETP},
	pages = {762-770},
	title = {Nonuniform state of superconductors},
	volume = {20},
	year = {1965}}

\end{document}